\documentclass[%amsmath,amssymb
    ,final           % use final for the camera ready runs
    ,numberedheadings % uncomment this option for numbered sections
  ,cmfonts,sorted&compressed]
  {aipproc}
\usepackage{amsmath}

\layoutstyle{8x11single}
\newcommand{\sig}[1]{{\boldsymbol \sigma}_{#1}}

\newcommand{\be}{\begin{equation} }
\newcommand{\ee}{\end{equation}}

\newcommand{\bea}{\begin{eqnarray}}
\newcommand{\eea}{\end{eqnarray}}

\newcommand{\secl}[1]{ \label{sec:#1}}

\newcommand{\eqnl}[1]{\label{eqn:#1}}

\newcommand{\figc}[1]{Figure \ref{fig:#1}}

\newcommand{\tot}{{\rm tot}}

\newcommand{\kd}{k_{\rm d}}
\newcommand{\bn}{\hat{\bf n}}
\newcommand{\bx}{{\bf x}}
\newcommand{\bk}{{\bf k}}

\newcommand{\wm}{\Omega_m h^2}
\newcommand{\wb}{\Omega_b h^2}

\newcommand{\bl}{{\bf l}}

\def\spose#1{\hbox to 0pt{#1\hss}}
\def\simlt{\mathrel{\spose{\lower 3pt\hbox{$\mathchar"218$}}
     \raise 2.0pt\hbox{$\mathchar"13C$}}}
\def\simgt{\mathrel{\spose{\lower 3pt\hbox{$\mathchar"218$}}
     \raise 2.0pt\hbox{$\mathchar"13E$}}}
\def\simpropto{\mathrel{\spose{\lower 3pt\hbox{$\mathchar"218$}}
     \raise 2.0pt\hbox{$\propto$}}}

\newcommand{\ApJL}{Astrophys. J Lett.}
\newcommand{\AsAs}{Astron. Astrophys.}
\newcommand{\ApJ}{Astrophys. J}
\newcommand{\AJ}{Astron. J}
\newcommand{\PRL}{Phys. Rev. Lett.}
\newcommand{\PRD}{Phys. Rev. D}
\newcommand{\MNRAS}{Mon. Not. R. Ast. Soc.}
\newcommand{\ARAA}{Ann. Rev. Astron. Astrophys.}

\newcommand{\amp}{and}
\newcommand{\etal}{{\it et al.~}}
\newcommand{\aut}[2]{{#2.\ #1,}}
\newcommand{\laut}[2]{{#2.\ #1,}}
\newcommand{\refs}[6]{{\it #2} {\bf #3} (#5), {#4}.}

\newcommand{\mybib}[2]{\bibitem{#2}}

\input epsf
\begin{document}

\title{CMB Temperature and Polarization Anisotropy Fundamentals}

\author{Wayne Hu}{address={Center for Cosmological Physics and Department of Astronomy
	 and Astrophysics, University of Chicago, Chicago IL 60637}}
	
\begin{abstract}
The tremendous experimental progress in cosmic microwave background (CMB)
 temperature and polarization 
anisotropy studies over
the last few years has helped establish a standard paradigm for 
cosmology at intermediate epochs and has simultaneously raised
questions regarding the physical processes at the two opposite ends of time.
What is the physics behind the source of structure in the
universe and the dark energy that is currently accelerating its expansion?
We review the acoustic phenomenology that forms the cornerstone of
the standard cosmological model and discuss internal consistency relations
which lend credence to its interpretation.  We discuss future milestones
in the study of CMB anisotropy that have implications for inflationary and
dark energy models.  These include signatures of gravitational waves
and gravitational lensing in the polarization fields and secondary
temperature anisotropy from the transit of CMB photons across the
large-scale structure of the universe.
\end{abstract}

\maketitle
\section{Introduction}

The pace of discovery in the field of Cosmic Microwave Background (CMB) 
anisotropy has been accelerating over the last few years.  With it,
the basic elements of the cosmological model have been falling into place: the nature
of the initial seed fluctuations that through gravitational instability 
generated all of the structure in the universe, and the mixture of
matter-energy constituents that drives its expansion.  Since even reviews
written a scant year ago (e.g. \cite{HuDod02,WhiCoh02,BerMaiMen02}) 
seem out of date today,
we will focus mainly on
the fundamental physical elements that enter into
cosmic microwave background observables.

In 1992, the COBE DMR experiment reported the first detection of cosmological 
anisotropy in the temperature of the CMB \cite{Smoetal92}.  The $10^{-5}$ variations
in temperature detected on scales larger than the $7^\circ$ resolution 
provided strong support for the gravitational instability paradigm. These variations
represent the direct imprint of initial gravitational potential perturbations through
their redshifting effect on the CMB photons, called the Sachs-Wolfe effect
\cite{SacWol67}, and are of the right amplitude to explain the large-scale structure
of the universe.
From 1992-1998, a host of experiments (see references in \cite{BonJafKno00})
detected a rise and fall in the level of anisotropy from degree scales to 
arcminute scales. In the following two years, the Toco \cite{Miletal00},
Boomerang \cite{deBetal00}, and Maxima \cite{Hanetal00} experiments measured
a clearly defined first peak in the spectrum and provided empirical evidence
that the small scale anisotropy is dominated by coherent acoustic phenomena,
long predicted to exist \cite{PeeYu70,DorZelSun78}.  
This first peak measurement has subsequently been confirmed by several
groups, with the best measurement to date from the 
Archeops experiment \cite{Benetal02}.
These measurements firmly establish initial gravitational potential or 
curvature fluctuations as the primary source of structure in the
universe 
(c.f. cosmological defect models \cite{AlbCouFerMag96,Alletal97,SelPenTur97}),
provide clear evidence that the universe is close to 
spatially flat and, in conjunction with other cosmological measurements 
of the dark matter, an indicate that there is a component of 
missing or dark energy.  
These findings provide support for the inference
from distant supernovae that the expansion is currently accelerating
under the influence of the dark energy \cite{Rieetal98, Peretal99}.

In 2001, the DASI and Boomerang \cite{Haletal02,deBetal02} experiments
announced a detection of secondary acoustic peaks in the spectrum which was 
subsequently confirmed by the VSA experiment \cite{Scoetal02}.  These experiments
provided a precise measurement of the baryon-photon ratio that is in
excellent agreement with inferences from big bang nucleosynthesis 
(e.g. \cite{SchTur98}), lending
further confidence in the underlying cosmological model.  Moreover they
provided the first direct evidence that dark, non-baryonic, matter exists
in the early universe.  In the past year, two more predicted phenomena have
been discovered: the dissipation of the acoustic peaks at small scales
by the CBI experiment \cite{Peaetal02} and, following a series of 
increasingly tighter upper limits \cite{Keaetal01,Hedetal02},
the polarization of the anisotropy by the DASI experiment \cite{Kovetal02}.
These observed phenomena provide the best internal evidence that the physical
assumptions underlying the interpretation of the peaks are justified.
 
The past few years have been an era of milestones for experiments 
and millstones for theoretical speculation.  The
standard cosmological paradigm of structure formation by 
gravitational instability of cold dark
matter in a nearly homogeneous and isotropic universe 
has been so thoroughly tested that 
it is extremely difficult
to find viable contenders that differ in any fundamental way.  
The acoustic peaks are furthermore strikingly consistent
with two fundamental predictions of simple inflationary models of
the early universe: a flat spatial geometry and a nearly scale-invariant
spectrum of initial curvature fluctuations. 
Alternatives to inflation, must now be nearly indistinguishable 
phenomenologically to be consistent with the data (e.g. \cite{SteTur01}).

The future may see a reversal of this trend as CMB experiments
and theory focus on mysteries at the two opposite ends of time.
The detailed physics of inflation (or contending theories) 
and the dark energy remains unexplained
and largely unexplored.  
Here the increase in precision, and perhaps more importantly accuracy,
expected from future experiments will allow studies of the associated
phenomena: 
the statistical
properties of the anisotropy, 
features in the initial curvature spectrum, 
subtle effects in the polarization, 
and the secondary anisotropy generated by 
structure in the universe.  In fact
secondary anisotropy from unresolved clusters of galaxies may have 
been detected recently by the CBI \cite{Masetal02} and 
BIMA \cite{Dawetal02} experiments.

Key past and future milestones are summarized in Table 1.
In this review we shall try to place this somewhat bewildering 
array of phenomena in its cosmological context.  We begin with 
a description of CMB observables in \S
\ref{sec:observables}, proceed through the physical basis of
their generation in \S \ref{sec:plasma} to the phenomenology
of the acoustic peaks and secondary anisotropy in \S \ref{sec:acoustic}-\ref{sec:secondary}.  Finally in \S \ref{sec:discussion} we outline
future milestones of CMB temperature and polarization anisotropy.

\begin{table}
\begin{tabular}{lll}
\hline
  \tablehead{1}{l}{b}{Phenomena}
  & \tablehead{1}{l}{b}{Experiments}
  & \tablehead{1}{l}{b}{Date} \\
\hline
Sachs-Wolfe & COBE DMR & '92 \\
Degree-scale & many & '93-99 \\
First peak & Toco, Boom, Maxima & '99-'00 \\
Secondary peaks & DASI, Boom & '01 \\
Damping tail & CBI & '02 \\
Polarization detection & DASI & '02 \\
Secondary anisotropy & CBI?, BIMA? & '02? \\
Polarization peaks & -- & future \\
non/Gaussianity & -- & future \\
Reionization bump & -- & future\\
Lensing of peaks & -- & future \\
Dark energy ISW & -- & future \\
Grav.-wave polarization &  -- & future 
\end{tabular}
\caption{CMB anisotropy milestones}
\label{tab:a}
\end{table}

%%%%%%%%%%%%%%%%%%%%%%%%%%%%%%%%%%%%%%%%%%%%
%% MAINMATTER
%%%%%%%%%%%%%%%%%%%%%%%%%%%%%%%%%%%%%%%%%%%%

\section{CMB Observables}
\label{sec:observables}
The fundamental observable in the CMB is the intensity of radiation per 
unit frequency 
per polarization at each point in the sky.
The polarization state of the radiation in a direction of sky denoted $\bn$ is described by the intensity
matrix $\left< E_i(\bn) E_j^*(\bn) \right>$ where ${\bf E}$ is the electric field
vector and the brackets denote time averaging.  

The radiation is measured to be a blackbody of $T=2.728 \pm 0.004$K 
(95\% CL) \cite{Fixetal96} with fractional variations across the sky 
at the $10^{-5}$ level
\cite{Smoetal92} and fractional polarization at the $10^{-6}$ level \cite{Kovetal02}.
It is then convenient to describe the observables by a temperature fluctuation
matrix decomposed in the Pauli basis (e.g. \cite{BonEfs87,Kos96})
\begin{align}
{\bf P} &= C \left< {\bf E}(\bn) \, {\bf E}^\dagger(\bn) \right>  \nonumber \\
        &= \Theta(\bn) {\bf I}
        + Q(\bn)  \, \sig{3}
        + U(\bn) \,  \sig{1}
        + V(\bn) \,  \sig{2}
\,,
\end{align}
where we have chosen the constant of proportionality so that the Stokes parameters
($\Theta$,$Q$,$U$,$V$) are dimensionless, e.g. $\Theta(\bn) \equiv 
\Delta T(\bn)/T$ averaged over polarization states. Note that the
circular polarization $V$ is absent cosmologically and under 
a counterclockwise rotation of the coordinate axes by $\psi$,
$Q \pm i U \rightarrow e^{\mp 2 i \psi} (Q \pm i U)$.

The temperature and polarization fields are decomposed as 
\cite{KamKosSte97,ZalSel97}
\begin{align}
 \Theta_{l m} & = \int d{\bn}\, Y_{l m}^*(\bn)
 \Theta(\bn)\,, \nonumber\\
 E_{l m} \pm  i B_{l m} &= -\int d\bn {}_{\pm 2}\, Y_{l m}^*(\bn)
[Q(\bn) \pm i U(\bn)]\,,
\eqnl{thdecompose}
\end{align}
in terms of the complete and orthogonal set of spin harmonic functions, ${}_s Y_{l m}$,
which are eigenfunctions of the Laplace operator on
a rank $s$ tensor \cite{NewPen66,Goletal66}.  $Y_{l m} = {}_0 Y_{l m}$ is the ordinary spherical 
harmonic.  For small sections of sky, the spin-harmonic expansion becomes
a Fourier expansion with $Y_{l m} \rightarrow e^{i \bl \cdot \bn}$ 
and ${}_{\pm 2} Y_{l m} \rightarrow -e^{\pm 2i\phi_l} e^{i \bl \cdot \bn}$,
where $\phi_l$ is the azimuthal angle of the Fouier wavevector $\bl$.  Note
that the $E$ and $B$ modes are then simply the $Q$ and $U$ states in
the coordinate system defined by $\bl$ \cite{Sel97}, i.e. $B$-modes
have a polarization orientation at $45^\circ$ to the wavevector.  This decomposition
is analogous to the vector field case where the divergence-free and curl-free
portions are distinguished by the orientation of the velocity 
vector with respect to the 
wavevector.  Represented as $E(\bn) = \sum E_{l m} Y_{l m}(\bn)$ 
and $B(\bn) = \sum B_{l m} Y_{l m}(\bn)$, they describe scalar and 
pseudoscalar fields on the sky and hence are also distinguished by parity.

\begin{figure}[t]
\centerline{\epsfxsize=6.5in\epsffile{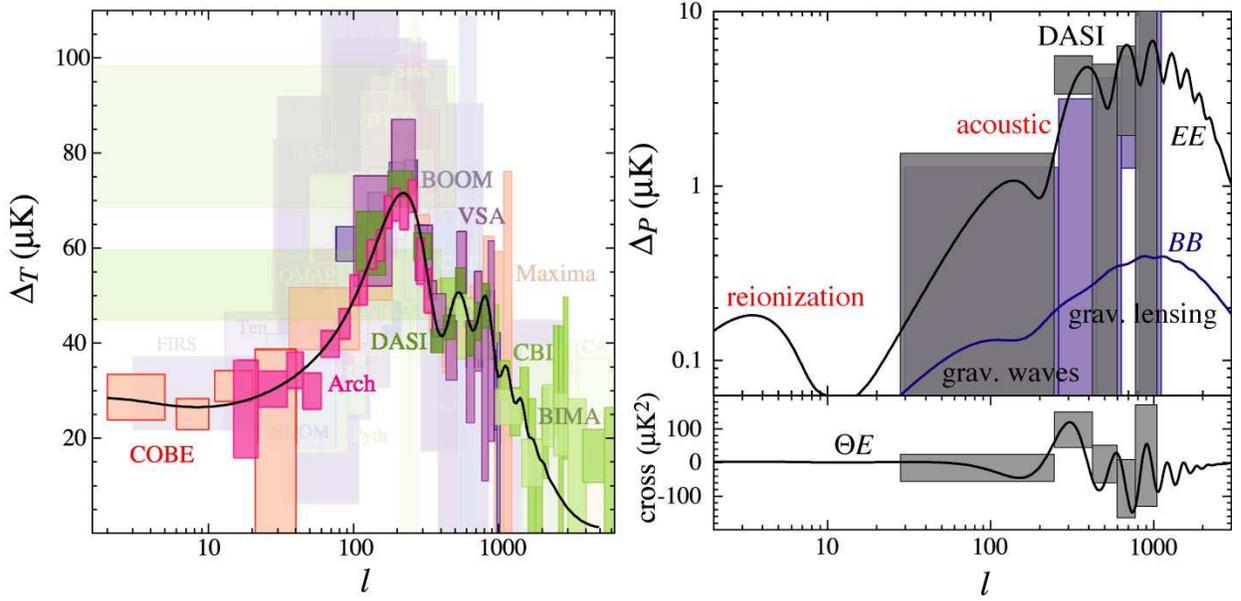}}
%\centerline{\epsfxsize=6.5in\epsffile{testfig.ps}}
\caption{\footnotesize Power spectra data plotted as the rms contribution 
per logarithmic interval $[l(l+1) C_l/2\pi]^{1/2}$ with error boxes
representing $1 \sigma$ error bars and approximate multipole bandwidth.  
Overplotted is a scale-invariant, 
flat cosmological model with $\Omega_m = 1/3$, $\Omega_\Lambda=2/3$, 
$h=0.7$, $\Omega_b h^2 = 0.02$,
reionization redshift $z_i=7$, and an inflationary energy 
scale of $E_i = 2.2 \times 10^{16}$ GeV.  
}
\label{fig:data}
\end{figure}

The primary observable is the two point correlation between fields 
$X, X' \in \{ \Theta, E, B \}$
\begin{align}
    \left< X_{l m}^{*} X_{l'm'}' \right> & =
   \delta_{l l'}\delta_{m m'} C_{l}^{XX'}\,,
\eqnl{cldef}
\end{align}
and are described by power spectra $C_l$ as long as the fields are 
statistically
isotropic.  If parity is also conserved then $B$ has no correlation with $\Theta$ or  $E$.
If in addition the fluctuations are Gaussian distributed,
the power spectra contain all of the statistical information about the
fields.  The measurements of these power spectra to date are shown in
Fig.~\ref{fig:data}.

\section{Photon-Baryon Plasma}
\label{sec:plasma}

\subsection{Cosmological Background}

We begin with a brief review of the cosmological background and the 
parameters that govern it.  
The expansion of the universe is described by the scale
factor $a(t)$, set to unity today, or equivalently 
the redshift $z(t) = a(t)^{-1}-1$, 
and by the current expansion rate,
the Hubble constant $H_0=100h$ km sec$^{-1}$ Mpc$^{-1}$.
The universe is {\it flat} (no spatial curvature) if the total
density is equal to the critical density,
$\rho_c=1.88h^2\times 10^{-29}$g cm$^{-3}$; it is {\it open} (negative
curvature) if the density is less than this and {\it closed} (positive
curvature) if greater.
The mean densities of different components of the
universe control $a(t)$ and are typically expressed today in units
of the critical density $\Omega_i$, with an evolution $\rho_i \propto 
a^{-3(1+w_i)}$, where $w_i=p_i/\rho_i$ with 
$p_i$ is the pressure.  In particular the
CMB photons have $w_\gamma=1/3$ and so $\rho_\gamma \propto a^{-4}$ or
$T \propto a^{-1}$. 
The quantities $\Omega_i h^2$ are proportional to the physical density
of the species today.  We will be interested in the baryonic
component $\Omega_b h^2$ and the total non-relativistic matter $\Omega_m h^2$,
where the difference is made up of cold dark matter.  We also allow for
a dark energy component $\Omega_e$, with an equation of state parameter $w_e$
which for illustrative purposes we often 
take to be
a cosmological constant ($\Omega_\Lambda$, $w_\Lambda=-1$) 
as consistent 
with indications from distant supernovae \cite{Rieetal98,Peretal99}.
We represent the total density as $\Omega_\tot = \sum_i \Omega_i$ so that
deviations from unity represent a spatially curved universe.

\begin{figure}[t]
\centerline{\epsfxsize=5.0in\epsffile{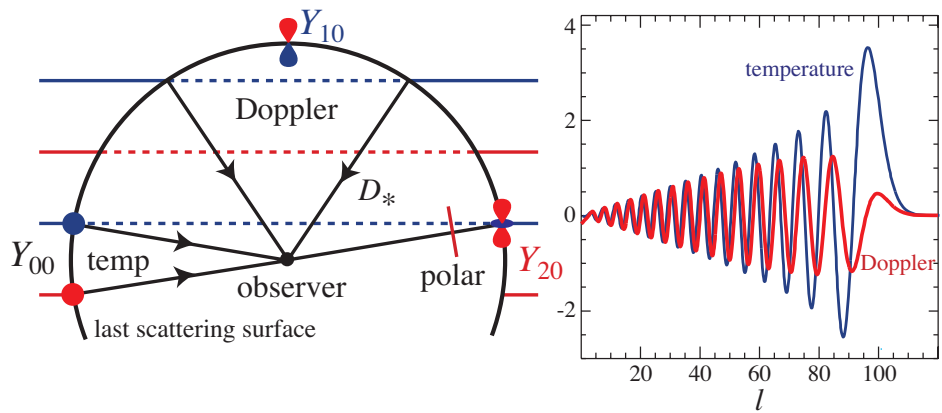}}
\caption{\footnotesize Geometric projection of sources at the last
scattering surface into temperature and polarization anisotropy today.  Left:
the main sources of anisotropy are the effective local temperature which
has a monopole structure ($Y_{00}$), the Doppler effect which has a 
dipole structure ($Y_{10}$) and the quadrupole source of polarization
($Y_{20}$) decomposed into plane waves.  Right: the projection of the
sources involves a coupling of the plane wave angular dependence or
momenta with that of the source.  A plane wave temperature 
inhomogeneity produces a sharp peak in the
anisotropy at $l \approx k D_* (=100$ here) 
reflecting the structure of $j_l(k D_*)$
shown;
the Doppler effect does not produce peaks since the coupling involves
$l \pm 1$ in the combination $j_l'(k D_*)$.  The functions here 
are weighted by $2l+1$ to account for the number of modes.} 
\label{fig:bessel}
\end{figure}

\subsection{Sources of Anisotropy}

As the CMB cools due to the expansion, the photons eventually no longer have
sufficient energy to overcome the
binding energy of hydrogen $B=13.6$eV to keep the medium ionized.
After this epoch, CMB photons propagate essentially
unimpeded to the observer.  This transition, called recombination,
occurs relatively rapidly around $a_* \approx 10^{-3}$ at an energy 
scale of $\sim 1/3$ eV as can be seen by the equilibrium
Saha equation
\begin{align}
{x_e^2 \over 1-x_e}
&= {n_e n_p \over n_H n_b} = {1 \over n_b} \left( { m_e T \over 2\pi }
\right)^{3/2} e^{-B/T}\nonumber\\
&\approx 3 \times 10^{15} \left( {\Omega_b h^2 \over 0.02} \right)^{-1}
	 \left( {B \over T} \right)^{3/2} e^{-B/T}\,,
\end{align}
where $x_e$ is the ionization fraction of hydrogen and we have neglected a
small contribution from helium.  The low $T/B$ of the
transition is mainly due to the low baryon-photon ratio of the universe.
It is sufficient that photons in the tail of the blackbody 
distribution be energetic enough to ionize hydrogen.
In reality, recombination is slightly less rapid than implied by the
Saha equation since processes which account for net recombination are not
sufficiently rapid to maintain equilibrium toward the end of recombination 
\cite{Pee68,ZelKurSun69} (see \cite{SeaSasSco00} for refinements).
Nonetheless the transition is sufficiently rapid 
that the photons in the CMB originate from a sharply defined surface 
or photosphere called the last scattering surface.

The properties of the plasma on this surface directly translate into the
observable primary temperature and polarization anisotropy in the CMB.
These properties are themselves
governed by Thomson scattering of photons off
of free electrons, which has a differential cross section of
\begin{equation}
{d \sigma \over d\Omega} = {3 \over 8\pi} |\hat {\bf E}' \cdot \hat{\bf E}|^2\sigma_T
\,,
\label{eqn:Thomson}
\end{equation}
where $\sigma_T = 8\pi \alpha^2/3m_e$ is the Thomson cross section,
$\hat {\bf E}'$ and $\hat {\bf E}$ denote the incoming and outgoing
directions of the electric field or polarization vector.
Summed over polarization and incoming angle, the comoving mean free path 
of a photon in the electrons is
\begin{align}
\dot\tau^{-1} &= {1 \over n_e \sigma_T a} \approx 2.5 \left( x_e 
{\Omega_b h^2 \over 0.02}
\right)^{-1} \left( {a_* \over 10^{-3}} \right)^2 {\rm Mpc}\,,
\label{eqn:mfp}
\end{align}
which is relatively small by cosmological standards.  Overdots here and
below represent derivatives with respect to conformal time. 
Electrons are coupled to baryons
by Coulomb interactions and so on scales larger than $\dot\tau^{-1}$,
the photon-baryon plasma can be considered a nearly perfect fluid \cite{PeeYu70}.  
In particular, rapid scattering keeps the photons 
isotropic in the baryon rest frame and so 
sets its dipole moment equal to the baryon velocity.

The tight coupling approximation implies that the CMB is
described by relatively few quantities.  There is the local temperature, or monopole,
 in space $\Theta_{00}({\bf x})$, the dipole $\Theta_{1m}({\bf x})$ and a residual but 
important
quadrupole moment $\Theta_{2m}({\bf x})$ from the breakdown of tight coupling.
In addition, the photons experience a gravitational redshift from gravitational
potentials and gravitational waves.  Modern CMB codes 
\cite{SelZal96,LewChaLas00} track the evolution of these sources 
\cite{BonEfs84,VitSil84} and construct the
present day CMB temperature and polarization anisotropy spectrum by 
geometric projection.

As a simple example, let us consider the spatial 
temperature perturbation $\Theta({\bf x}) \equiv \sqrt{4\pi} \Theta_{00}({\bf x})$, 
where the
normalization convention reflects that of $Y_{l m}$,  on a
last scattering surface considered to be infinitely sharp at a comoving distance $D_*$ (see Fig.~\ref{fig:bessel}).
In a flat universe, the spatial field may be represented by its Fourier harmonics
\begin{align}
\Theta(\bn) &= \int dD \Theta(\bx) \delta (D-D_*) \nonumber\\
                  &= \int {d^3 k \over (2\pi)^3} \Theta(\bk) e^{i {\bk \cdot D_* \bn}}\,.
\end{align}
In an spatially curved universe, the plane waves must be replaced by the eigenfunctions of
the Laplace operator to account for a change in the relationship between
distance and angle that we discuss below.  
Also to achieve high precision in the predictions, the
delta function must be replaced by the visibility function $\dot \tau e^{-\tau}$,
the probability of a photon last scattering in a distance interval $dD$
to account for the finite duration of recombination.

Expanding the plane waves in spherical harmonics, we obtain
\begin{align}
\Theta_{l m} = \int {d^3 k \over (2\pi)^3 }\Theta(\bk) 4\pi i^l j_l(k D_*) Y_{l m}(\hat \bk)\,.
\label{eqn:monopoleproj}
\end{align}
The power spectrum of the anisotropy contributions 
from the local temperature is then governed by that of the
spatial temperature field at recombination
\begin{equation}
\langle \Theta^*(\bk) \Theta (\bk') \rangle \equiv (2\pi)^3 \delta(\bk - \bk') P_T(k)\,,
\end{equation}
such that
\begin{align}
C_l^{\Theta\Theta} &= 4\pi \int {dk \over k}
\, j_l^2(k D_*) {k^3 P_T(k) \over 2\pi^2} \,.
\label{eqn:clpower}
\end{align}
Note that the quantity $k^3 P_T/2\pi^2$ is the contribution per log interval to the 
variance of the temperature $\sigma^2 = \int d^3 k/(2\pi)^3 P_T$.
For a slowly-varying log power, the integral in Eqn.~(\ref{eqn:clpower})
can be performed analytically
\begin{align}
{l(l+1) \over 2\pi}C_l^{\Theta\Theta} & \equiv \left( \Delta_T^2 \over T^2 \right) \approx 
{k^3 P_T(k) \over 2 \pi^2} \Big|_{k= l/D_*}\,,
\label{eqn:monopower}
\end{align} 
and so it is convenient to represent the anisotropy by the rms temperature
contribution per log interval $\Delta_T$.  
It is also the contribution to the anisotropy variance per log interval in $l$
for $l \gg 1$.     We also define the analogous quantities to describe 
the polarization fields ($\Delta_P$) and the temperature-polarization cross correlations
($l(l+1)C_l^{\Theta E}/2\pi$). 

This behavior of the local temperature is representative of 
the other sources with a few important differences.  
In general, power from low order multipole moments $l = 0,1,2$
in the CMB project into higher order anisotropy today by simple geometry
\cite{BonEfs87}.
The local temperature is an $l = 0$ source.
Note that the spherical Bessel function $j_l(x)$ peaks at $l \sim x$ (see Fig.~\ref{fig:bessel}
right) and
so the translation in Eqn.~(\ref{eqn:monopoleproj}) 
from power in real space to angular space is nearly one-to-one.
Thus in Eqn.~(\ref{eqn:monopower}) one takes $k D_* \approx l$ 
with features in one preserved in the other.    This corresponds to an edge-on
($\bn \perp \bk$)
projection of the wavelength of the fluctuation onto angle: $\theta = \lambda/D_*$. 

Other sources
of anisotropy have angular dependence on the last scattering surface which 
changes the relationship between $l$ and $k$.
The dipole term provides an instructive example.    It represents Doppler shifts in the temperature
due to the motion of the photon-baryon plasma along the line of sight.  For a curl free flow
as expected cosmologically,  there is no contribution to the Doppler effect for $\bn \perp \bk$
and the peak structure is washed out.  
Mathematically, the Doppler effect contributes a dipole
term $\bn \cdot \bk \propto Y_{1m}$ 
which when coupled to the plane wave angular momentum $Y_{l m}$
leads to a recoupling of $j_{l \pm 1}$ to $j_l'$ 
(see Fig.~\ref{fig:bessel}).  Unlike $j_l$, 
$j_l'$ has no peak structure 
leading to a one-to-many projection of $k$ onto $l$.
Geometrically, well-defined features such as peaks in the anisotropy spectrum cannot
come from the Doppler
effect \cite{HuSug95a}.   All sources may be described in this framework of the addition 
of local angular momentum with plane wave orbital angular momentum with the
net effect of modifying the mapping between $k$ and $l$ space \cite{HuWhi97c}.

The local quadrupole plays a special role in that it is the fundamental source of
linear polarization in the CMB as can be seen from the Thomson differential
cross section of Eqn.~(\ref{eqn:Thomson}). 
Heuristically, incoming radiation shakes an electron in the direction
of its electric field vector $\hat{\bf E}'$
causing it to radiate with an outgoing polarization parallel to that
direction.  However since the outgoing polarization
$\hat{\bf E}$
must be orthogonal to the outgoing direction, incoming radiation that
is polarized parallel to the outgoing direction cannot scatter, leaving
only one polarization state.  If the intensity were completely isotropic
the missing polarization state is supplied by radiation incoming from
the direction orthogonal to the original one.  
Only a quadrupole temperature anisotropy in the radiation
generates a net linear polarization from Thomson scattering.  In particular,
the polarization is oriented along the cold axis of the quadrupole moment 
in the transverse plane.
We shall see that the breakdown of tight coupling during recombination allows
quadrupoles to be generated out of gradients in the fluid velocity and from 
gravitational redshifts associated with gravitational waves.  For the former,
since the quadrupole axis is aligned with the wavevector, the polarization 
generated is a pure $E$-mode (see Fig.~\ref{fig:bessel}).  
Gravitational waves generate both $E$ and
$B$ modes \cite{KamKosSte97,ZalSel97}.

To understand the phenomenology of the anisotropy in the CMB temperature
and polarization fields it is sufficient to understand the dynamics governing the local sources: the monopole, dipole and quadrupole of the CMB temperature field and their gravitational or metric fluctuation 
analogues.

\subsection{Fluid Dynamics}

The evolution of the anisotropy sources is governed by simple fluid dynamics.
From the previous section, the critical variables are the monopole
or temperature fluctuation $\Theta(\bk) \equiv \sqrt{4\pi} \, \Theta_{00}(\bk)$, the
dipole or bulk velocity $v_\gamma(\bk) = -i \sqrt{4\pi/3}\,  \Theta_{10}(\bk)$, and
the quadrupole or anisotropic stress $\pi_\gamma(\bk) = -(12/5)\sqrt{4\pi/5} \, \Theta_{20}(\bk)$.
We will hereafter drop the argument $\bk$ with the understanding that the Fourier
representation is always assumed.
For convenience we have chosen the coordinate system so that ${\bf z} \parallel \bk$
so that the plane waves are azimuthally symmetric and stimulate only the $m=0$ mode (see Fig.~\ref{fig:bessel}).
Likewise we have suppressed the vector dependence of the bulk velocity  by the
same assumption ${\bf v}_\gamma = -i v_\gamma \hat \bk$.
The analogous quantities for the baryons are the density perturbation $\delta_{b}$ and bulk
velocity $v_{b}$.   For gravity, we choose a conformal Newtonian representation 
(see e.g. \cite{MaBer95}) where
the gravitational potential perturbations are defined by 
the Newtonian potential $\Psi$
(time-time metric fluctuation) and the curvature fluctuation
$\Phi$ (space-space metric fluctuation $\approx -\Psi$).

Covariant conservation of energy
and momentum requires that the photons and baryons satisfy separate continuity
equations 
\begin{eqnarray}
\dot \Theta = -{k \over 3} v_{\gamma} - \dot\Phi \, , \qquad
\dot \delta_b = -k v_b - 3\dot\Phi \, ,
\label{eqn:continuity}
\end{eqnarray}
and coupled Euler equations
\begin{align}
\dot v_{\gamma} &= k(\Theta + \Psi) - {k \over 6} \pi_\gamma
        - \dot\tau (v_\gamma - v_b) \, , \nonumber\\
\dot v_b &=- {\dot a \over a} v_b + k\Psi + \dot\tau(v_{\gamma} - v_b)/R
        \, ,
\label{eqn:Euler}
\end{align}
where $R= (p_b + \rho_b)/(p_\gamma + \rho_\gamma) \approx
3\rho_b / 4 \rho_\gamma$ is the photon-baryon momentum density ratio.
We have neglected a small correction to the anisotropic
stress term in a curved universe.

The continuity equations represent particle number conservation.
For the baryons, $\rho_b \propto n_b$.
For the photons, $T \propto n_\gamma^{1/3}$, which explains the
$1/3$ in the velocity divergence term.
The $\dot\Phi$ terms come from the fact that $\Phi$ is a perturbation to
the scale factor and so they are the perturbative 
analogues of the cosmological redshift 
and density dilution from the expansion.
The Euler equations have similar interpretations.
The expansion makes particle momenta
decay as $a^{-1}$.  The cosmological redshift of $T$ accounts for
this effect in the photons.
For the baryons, it becomes the expansion drag on
$v_b$ ($\dot a/a$ term).  Potential gradients $k\Psi$ generate
potential flow.  For the photons, stress gradients in the fluid, both
isotropic ($k\delta p_\gamma /(p_\gamma + \rho_\gamma) = k
\Theta$) and anisotropic ($k\pi_\gamma$) counter infall.
Thomson scattering
exchanges momentum between the two fluids
($\dot \tau$ terms).

For scales much larger than the mean free path $\dot\tau^{-1}$, the Euler equation may be
expanded to leading order in $k/\dot\tau$,
such that the photons are isotropic in the baryon rest frame $v_\gamma = v_b$
and so the joint Euler equation becomes
\begin{equation}
{d \over d\eta} [(1+R) v_\gamma] = k [\Theta + (1+R)\Psi]\,.
\end{equation}
Combining this with the continuity equation leads to the oscillator equation
(e.g. \cite{HuSug95a})
\begin{equation}
{d \over d\eta}[(1+R) \dot\Theta] + {k^2 \over 3}\Theta
= -{k^2 \over 3} (1+R)\Psi - {d\over d\eta} [(1+R)\dot\Phi] \,,
\label{eqn:oscillator}
\end{equation}
and a small residual anisotropic stress or quadrupole that tracks
the evolution of the fluid velocity \cite{Kai83}, 
\begin{equation}
\pi_\gamma = {32 \over 15} {k \over \dot \tau} v_\gamma\,.
\label{eqn:pitc}
\end{equation}
This dependence reflects the fact that a local quadrupole can arise
from a gradient in the velocity field, for example as photons from two hot
crests of a plane wave fluctuation meet at the trough in between (see
Fig.~\ref{fig:bessel} left), but is
suppressed by scattering.  

Equation (\ref{eqn:oscillator}) is the fundamental relation for
acoustic oscillations.  
The change in the momentum of the photon-baryon
fluid is determined by a competition between the
pressure restoring and gravitational driving forces which causes the system
to oscillate around its equilibrium.  
Note that the frequency of the oscillation
\begin{equation}
\omega^2 ={ 1\over 3(1+R) }  k^2 = c_s^2 k^2\,,
\end{equation} 
where $c_s$ is the sound speed of the fluid.

Equation (\ref{eqn:pitc})
is the fundamental relation for acoustic polarization.  Its quadrupole
source tracks the motion of the fluid.  Polarization is generally small, at most
$\sim$10\% of the anisotropy itself, since it requires 
scattering for its generation but 
its quadrupole source is suppressed if the scattering is too rapid.
Given the initial conditions and gravitational potentials,
these equations predict the phenomenology of the acoustic oscillations
in the temperature and polarization fields.

\subsection{Initial Conditions}

The simplest inflationary models 
make a set of definite predictions for the
initial conditions of the acoustic oscillations 
and hence their successful observation
provides strong support for the inflationary paradigm.  
Quantum fluctuations in the scalar field
that drives inflation imprints a nearly spectrum
of Gaussian curvature (potential) fluctuations $k^3 P_\Phi/2\pi^2 \propto
k^{n-1}$ where $n \approx 1$  
\cite{GutPi82,Haw82,BarSteTur83} on a spatially flat background metric.  
Gravitational infall into these
initial potentials eventually generates all of the
structure in the universe.
Quantum fluctuations in the gravitational wave degrees of freedom also produce a
nearly scale-invariant spectrum of fluctuations 
whose power depends on $E_i^4$
where $E_i$ is the energy scale of inflation \cite{RubSazVer82,FabPol83}. 

A Newtonian gravitational potential $\Psi \approx - \Phi$ necessarily imparts an initial temperature
perturbation since $\Psi$ represents a spatially varying time-time perturbation to the metric 
away from coordinates where the temperature is homogeneous.
The perturbation is equivalent to a change in the scale factor 
since 
\begin{equation}
a \propto t^{2 \over 3(1+p/\rho)}\,,
\end{equation}
which then produces a change in the temperature perturbation 
from the cosmological redshift
$T \propto a^{-1}$ of
\begin{equation}
\Theta = -{ 2 \over 3 (1+p/\rho)} \Psi
\end{equation}
or $-\Psi/2$ in the radiation dominated
era \cite{Pea91,WhiHu97}. We call $\Theta+\Psi$ the effective temperature since it
also accounts for the redshift a photon experiences when climbing out
of a potential well, also known as the Sachs-Wolfe effect \cite{SacWol67}.  
In the matter dominated epoch, $\Theta +\Psi  = \Psi /3$.

There are three important aspects of these results.  First, inflation sets
the {\it temporal} phase of all wavemodes by starting them all at
the initial epoch.  We shall see that this predicts a coherent 
set of peaks in the CMB spectrum with a definite phase.  Cosmological defect models 
predict a more random distribution of acoustic phases which produces incoherent acoustic phenomena
\cite{AlbCouFerMag96,Alletal97,SelPenTur97}.
Defects can be now be ruled out as a primary mechanism for structure formation in 
the universe.  
More generally, without inflation or some other modification to the matter-radiation
dominated universe, curvature perturbations
cannot be generated outside the apparent horizon and so build up only by the causal motion of matter.  This
generally entails at least a delay in temporal phase of the oscillations which is
not observed.

Secondly, since the power spectrum
of the effective temperature is directly related to scale-invariant curvature fluctuations 
from inflation, Eqn.~(\ref{eqn:monopower}) implies the acoustic oscillations should also be approximately  scale invariant in amplitude in the tight coupling regime.  Observations
are in excellent agreement with this fundamental prediction with
tight constraints on the index $n= 0.94^{+0.11}_{-0.04}$ (\cite{Bonetal02},
see also \cite{KnoChrSko01,WanTegZal02,Peretal02}).
We will therefore base the
 discussion of the acoustic phenomenology on models
with nearly scale-invariant initial curvature fluctuations as predicted by inflation.

Finally, the scale-invariant gravitational wave background leads to a quadrupolar distortion
in the CMB temperature field just like its effect on a ring of test masses.  Because the 
quadrupole axes lie in the plane transverse to the wavevector, this quadrupole anisotropy
leads to $B$-mode polarization as the tight coupling approximation breaks down.  
A measurement of $B$-modes from gravitational waves would determine the energy
scale of inflation $E_i$ but its strong scaling with $E_i$ implies that it can only 
open a relatively small window between a few $10^{15}$ -- few $10^{16}$ GeV for
possible detection \cite{KnoSon02}.

\begin{figure}[t]
\centerline{\epsfxsize=4.5in\epsffile{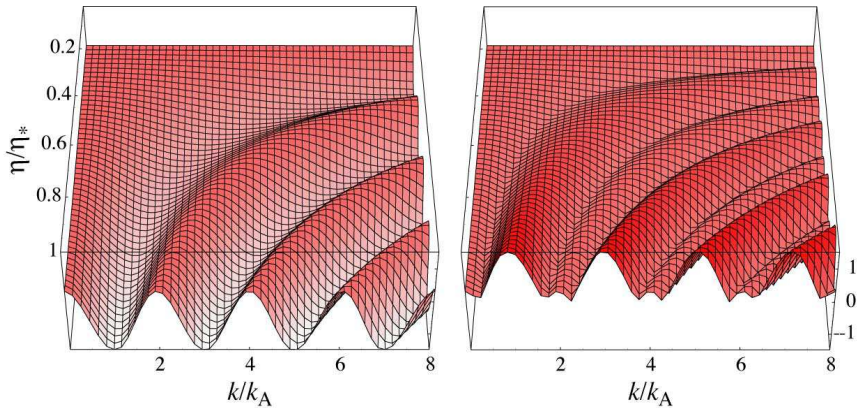}}
\caption{\footnotesize Temporal evolution of the effective temperature 
as a function 
of wavenumber predicted by the simplified model of Eqn.~(\ref{eqn:baryonoscillations}).  
Left: amplitude;
right: rms fluctuation.  All modes begin with the same phase and are frozen in
at the same time $\eta=\eta_*$.  The result is a harmonic series of extrema
with a fundamental scale of $k_A = \pi /s_*$.  Baryons, here shown with 
an $R = 0.6 (\eta/\eta_*)^{2}$, displace the zero point of the oscillation and enhance
the compressional troughs.  In the rms, this leads to enhanced odd numbered peaks.}
\label{fig:osc}
\end{figure}
 
\section{Acoustic Phenomenology}
\label{sec:acoustic}

\subsection{Acoustic Scale}

It is instructive to first consider a simplified model where the universe is 
always matter-dominated in its expansion and the dynamical effects
of the baryons are negligible.  This amounts to holding $\Phi$ and $\Psi$
constant in the oscillator equation (\ref{eqn:oscillator}) and setting
$R=0$.  The solution for an initial
curvature fluctuation is then simple
\begin{equation}
[\Theta + \Psi](k,\eta) = { 1\over 3} \Psi (k,0) \cos(k s)\,,
\qquad
v_\gamma(k,\eta) = {\sqrt{3} \over 3} \Psi(k,0) \sin(k s)\,,
\label{eqn:simpleoscillator}
\end{equation}
where $s=\int_0^{\eta} c_s d\eta'$ is the distance sound
can travel by $\eta$ or the {\it sound horizon}.

In the limit of scales large compared with the sound horizon $ks \ll 1$,
the perturbation is frozen into its initial conditions.  This is the gist of
the statement that the large-scale anisotropy measured by COBE directly
measure the initial conditions.  On small scales, the amplitude of the
Fourier modes
exhibits temporal oscillations corresponding to compression and
rarefaction of the plasma inside gravitational potential wells.
All modes are frozen in at recombination $\eta_*$ (see Fig.~\ref{fig:osc} left).
Modes that are caught at either maxima {\it or} minima of their oscillation at
recombination correspond to peaks in the rms fluctuation or power 
(see Fig.~\ref{fig:osc} right).  
Given the solution in Eqn.~(\ref{eqn:simpleoscillator}), these modes
are harmonics 
$k_n = n k_A$ 
of a fundamental acoustic scale 
$k_A = \pi /s_*$, where $n$ is an integer (see \figc{osc}a).     
From Eqn.~(\ref{eqn:monopower}), these become angular peaks in the
anisotropy power with a characteristic angular scale of $l_A = k_A D_*$.
The Doppler effect from the line-of-sight velocity has an 
rms $v_\gamma/\sqrt{3}$ and so 
contributes an equal amplitude fluctuation.  It 
is $\pi/2$ out of phase with the
temperature.  This is because extrema represent turning points in the
oscillation where the velocity vanishes.  Nonetheless, because of the
geometry of the projection (see Fig.~\ref{fig:bessel}), these oscillations 
do not contribute peaks in
the angular power spectrum.   We shall see that they do however predict
the peaks in the polarization spectrum through Eqn.~(\ref{eqn:pitc}).

\begin{figure}[t]
\centerline{\epsfxsize=5.0in\epsffile{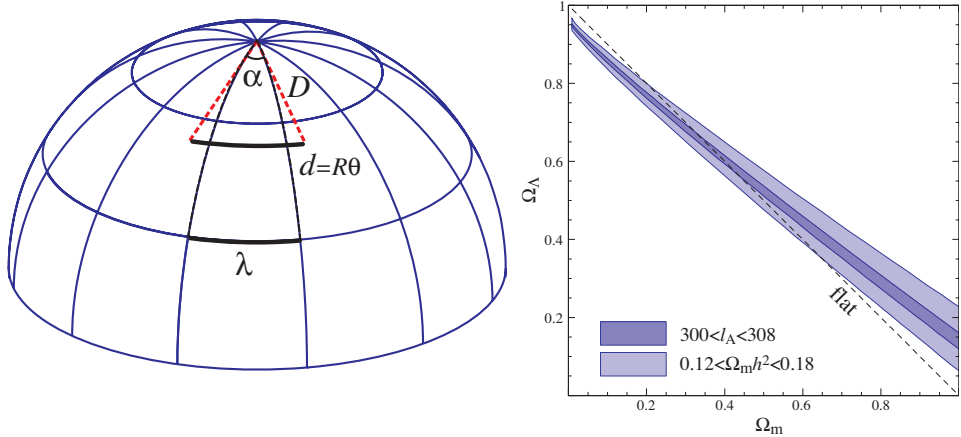}}
\caption{\footnotesize The acoustic scale.  Left: geometrical effect in a closed universe.
Objects in a closed universe are further than they appear ($D < d$).  Consequently the
measured angular scale of the peaks is extremely sensitive to the curvature, once
the physical scale has been set by the acoustic physics.
Right: translation of a constraint on the peak scale $300 < l_A < 308$ \cite{KnoChrSko01}
(68\% CL) onto the cosmological parameters $\Omega_m$ and $\Omega_\Lambda$ with all other
parameters fixed versus constraints
with uncertainties in the physical matter density $\Omega_m h^2$ from CMB determinations
\cite{Bonetal02} folded in.  Uncertainties in this plane are no longer dominated by peak 
location measurement errors.
}
\label{fig:omol}
\end{figure}

Under the flat, matter-dominated assumption the horizon distance
scales as $\eta \propto a^{1/2}$.  Thus $s_* \approx D_* /\sqrt{3000}$
and so $l_A \approx 200$. 
In a spatially curved universe, the distance used in the conversion from 
$k$ to $l$ no longer equals the comoving distance $D_* \ne \eta(a=1)-
\eta_*$.  
Consider first
a closed universe with radius of curvature
$R =  H_{0}^{-1}|\Omega_\tot-1|^{-1/2}$.
Suppressing one spatial coordinate yields
a 2-sphere geometry with the observer situated at the
pole (see Fig.~\ref{fig:omol} left).  Light travels on
lines of longitude.
A physical scale $\lambda$ at fixed latitude given by
the polar angle $\theta$ subtends an angle
$\alpha = \lambda/R\sin\theta$.
For $\alpha \ll 1$,
a Euclidean analysis would infer a
distance $D  =R\sin\theta$, even though
the {\it coordinate distance} along the arc is
$d = \theta R$; thus
\begin{equation}
D = R\sin( d / R)\,.
\end{equation}
For open universes, simply replace $\sin$ with $\sinh$.  
The peak locations $l_n = n k_A D_*$ making them extremely
sensitive to the spatial curvature 
\citep{DorZelSun78,KamSpeSug94};
measurements constrain the geometry to be close to flat, consistent
with a radius of curvature that is larger than the observable
universe.  Since a flat universe is at the critcal density, local 
measurements of a low matter density indicate
that there is a missing or ``dark" component of the energy density, 
in good agreement with
indications of an accelerating expansion from distant supernovae.

\begin{figure}[t]
\centerline{\epsfxsize=4.5in\epsffile{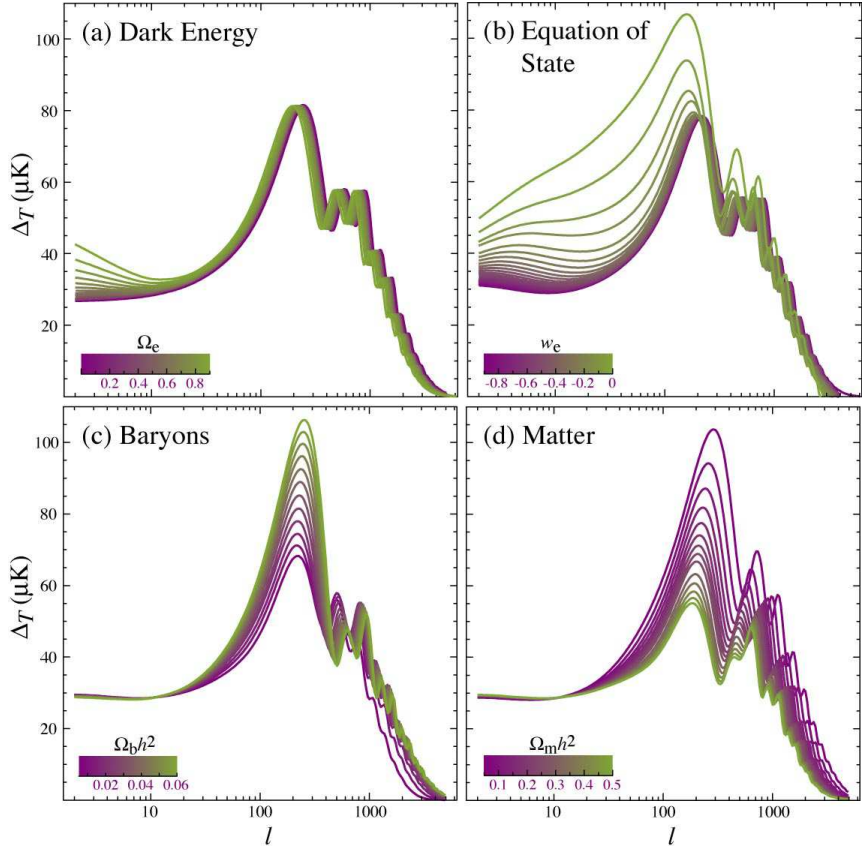}}
\caption{\footnotesize Sensitivity of the temperature power spectrum to four fundamental
cosmological parameters: the energy density of the dark energy today $\Omega_e$ in units of
the critical density, the equation of state parameter of the dark energy $w_e$, the physical baryon
density $\Omega_b h^2$ and the physical matter density $\Omega_m h^2$.  All are varied around
a fiducial flat model of $\Omega_e=0.65$, $w_e=-1$, $\Omega_b h^2 =0.02$ and $\Omega_m h^2=0.15$ with $n=1$.}
\label{fig:cl}
\end{figure}

These simple scalings must be modified for the fact that the universe is not
fully matter dominated at recombination, the fluid is not purely 
photon-dominated, and the expansion is dark energy dominated today.  
The main effect is from the radiation density which changes the expansion
rate and so shifts the acoustic scale to higher multipoles.
For a flat universe with $\Omega_m h^2 \approx 0.15$, $l_A \approx 300$
which is a substantial shift.  The baryons lower the sound speed
and so have a similar but smaller effect.
The dark energy also provides a smaller effect through a decrease in
the distance 
to recombination $D_*$ 
and hence a lowering of $l_A$,
with an increase in $\Omega_e$ or $w$.
The sensitivity of the acoustic scale to cosmological parameters is approximately (see also Fig.~\ref{fig:cl})
\begin{eqnarray}
    {\Delta l_A \over l_A}
    \approx
    -1.1  {\Delta \Omega_\tot \over \Omega_\tot} 
    -0.24 {\Delta \wm \over \wm}
    -0.17 {\Delta \Omega_{e} \over \Omega_{e}}
    -0.11 \Delta w_e
    +0.07 {\Delta \wb \over \wb}\,,
\end{eqnarray}
around a model of $\Omega_\tot=1$, $\wm= 0.15$$, \Omega_e = 0.65$,
$w_e =-1$ and $\wb=0.02$ \cite{HuFukZalTeg00}.  A joint analysis of the data 
yields $l_A = 304 \pm 4$ \cite{KnoChrSko01} and with these
errors, the uncertainty in the cosmological interpretation 
in the dark energy and curvature
domain is already dominated by uncertainty in $\Omega_m h^2$ not
peak measurement error (see Fig.~\ref{fig:omol} right).  
Fortunately as we shall see,  this parameter
can be internally 
determined from the CMB and serves as an example where future increased
precision will have important implications for a quantity of fundamental
interest, the dark energy. 

Finally, there is a separate effect on the first peak.
We shall see that the decay of the gravitational potential in a
radiation dominated universe generates anisotropy from 
gravitational redshifts after last scattering, filling
in the rise to the 
first peak and shifting its location downwards off
of the harmonic series to $l_1 \approx 3l_A/4$, placing $l_1
\approx 220$ in agreement with the observed location \cite{Benetal02}.  

\subsection{Baryon loading}

\secl{loading}

Baryons add to the mass of the photon-baryon plasma without adding
to the pressure.  An examination of Eqn.~(\ref{eqn:oscillator}) shows that 
their effect comes solely through the baryon-photon momentum
density ratio $R \approx 0.6 (\Omega_b h^2 /0.02) (a/10^{-3})$.
 It is instructive to look at the solution to the oscillator 
equation (\ref{eqn:simpleoscillator}) in the approximation that 
$R$ is constant.   Again under the assumption
of a matter-dominated universe, the solution is that of 
a simple harmonic oscillator in a constant gravitational field but with an
increased mass term
\begin{align}
[\Theta+ \Psi](k,\eta_*) &= [\Theta+(1+R)\Psi](k,0)\cos(ks)\nonumber\\
& = [1+3R]{1\over 3}\Psi(0) \cos(ks) - R\Psi(0)\,,
\label{eqn:baryonoscillations}
\end{align} 
where the sound speed entering into the sound horizon calculation
is reduced since $c_s^2 = 1/3(1+R)$.
Aside from this reduction,
baryons have two distinguishing effects:
they enhance the amplitude of the oscillations by $1+3R$ fractionally and
shift the zero point of the oscillation by  $-R\Psi$.
The latter modulates the amplitude of neighboring peaks: odd numbered peaks
will be enhanced over the zero-baryon case by $1+6R$; even numbered
peaks remain the
same (see Fig.~\ref{fig:osc}, right).  Physically,
the baryon mass enhances compression inside gravitational potential wells. 

These qualitative results 
remain true in the presence of the real time-variable $R$.  
Measurement of these baryonic signatures currently limits 
$\Omega_b h^2 = 0.022^{+0.004}_{-0.002}$ \cite{Bonetal02}.  This 
value is strikingly
consistent with inferences of the baryon density from big bang nucleosynthesis.
Consequently there have been no significant changes in the baryon-photon ratio
from an energy scale of an MeV to an eV in the expansion history of the
universe.

\subsection{Dark Matter and Radiation}
\secl{driving}

The matter-to-radiation ratio scales
as $\rho_m/\rho_r \approx 3.6 (\Omega_m h^2/0.15) (a/10^{-3})$
and so the universe is only barely matter-dominated at last scattering.
Moreover fluctuations corresponding to the higher peaks entered the
sound horizon at an earlier time, during radiation domination.
As we have seen, including the radiation changes the expansion rate of the universe
and hence the physical scale of the sound horizon at recombination. 
Fortunately, radiation has the more unique effect of
driving the acoustic oscillations 
by making the gravitational force evolve with time \citep{HuSug95a}. 

The exact 
evolution of the potentials is determined by the relativistic Poisson
equation.  Its qualitative content is clear: since 
the background density is decreasing with
time, the density fluctuations must grow unimpeded
by pressure to maintain constant potentials.  
In particular, in the radiation dominated era
once pressure begins to fight gravity at the first compressional
extrema of the oscillation, the Newtonian gravitational potential and
spatial curvature must decay. % (see \figc{driving}).  

\begin{figure}[t]
\centerline{\epsfxsize=3.5in\epsffile{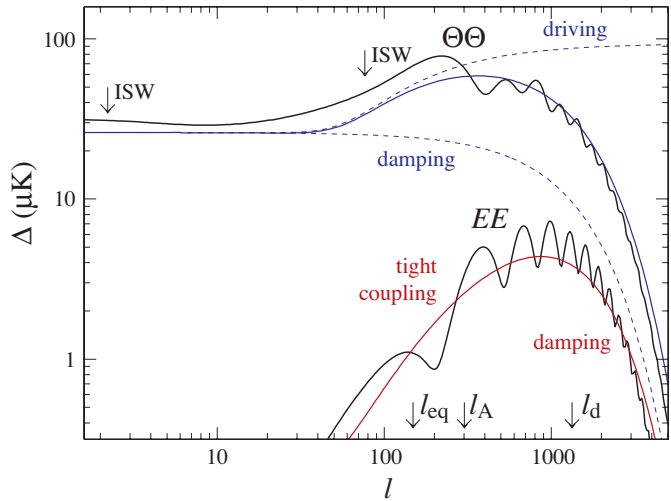}}
\caption{\footnotesize Phenomenological contributions to the anisotropy form.
Radiation drives the amplitude of the acoustic oscillations up across the
multipole scale subtended by the horizon at matter radiation equality $l_{\rm eq}$
and down across the damping scale $l_{D}$.  Functional forms for these 
envelopes are taken from \cite{HuWhi97a}.  The ISW effect pushes $l_1 < l_A$
due to the radiation and raises the large angle anisotropy due to the dark energy. 
The polarization scales as $l/l_D$ for $l < l_D$ due to the 
tight-coupling suppression of its quadrupole source and is also damped below $l_D$.}
\label{fig:scheme}
\end{figure}

This decay actually drives the
oscillations: it is timed to leave the fluid maximally compressed with
no gravitational potential to fight as it turns around.
The net effect is doubled since the redshifting from the spatial metric fluctuation
$\Phi$ also goes away at the same time.
When the universe becomes matter dominated
the gravitational potential is no longer determined by
photon-baryon density perturbations but by the pressureless
dark matter. Therefore, the amplitudes of the acoustic peaks
increase as the matter-to-radiation ratio decreases
\citep{Sel94,HuSug95a}.
The net result is that across 
the horizon scale at
matter-radiation equality the acoustic amplitude increases by a factor of
\begin{equation} 
{2\Psi  - \frac{1}{3} \Psi \over \frac{1}{3} \Psi } = 5
\end{equation}
for a pure photon and dark matter universe, and a factor of
4 when including the effect of neutrinos and baryons \cite{HuSug96} (see also
Fig.~\ref{fig:scheme}).
By eliminating gravitational potentials, radiation also
eliminates the alternating peak heights from baryon loading (see Fig.~\ref{fig:cl}).
The observed high third peak (see Fig.~\ref{fig:data}) is a good indication 
that matter dominates the energy density 
at recombination.  
Finally, the effect of the decaying 
potential after recombination 
leads to so-called integrated Sachs Wolfe contributions to 
the temperature fluctuations
through the continuity equation (\ref{eqn:continuity}) and shifts
the first acoustic peak downwards off of the acoustic series \cite{HuSug95a}. 

Observations of these matter-radiation phenomena (see Fig.~\ref{fig:cl}) 
currently constrain the
total matter density to be $\Omega_m h^2 = 0.15 \pm 0.03$ \cite{Bonetal02}
and are crucial in internally resolving ambiguities in the interpretation of the
peak scale.  Since this number greatly exceeds the measured baryon density, 
these observations are also the first empirical indication that the dark 
matter exists at high redshift.
 
\subsection{Damping}
\secl{damping}

The photon-baryon fluid has slight imperfections corresponding to
shear viscosity $\pi_\gamma$ and heat conduction $(v_b - v_\gamma)$ 
in the Euler equation (\ref{eqn:Euler}) \cite{Sil68,Wei71}.  These imperfections damp
acoustic oscillations.  These are both associated with the diffusion of the
photons which is especially important during recombination
when the mean free path of the photons rises dramatically.
As we have seen the viscosity $\pi_\gamma \sim k v_\gamma/\dot\tau$.
With the continuity equation
$k v_\gamma \approx -3 \dot \Theta$,
this leads to a $\dot \Theta$ damping term in the oscillator equation.
The heat conduction term can be shown to have a similar effect
by expanding the
Euler equations in $k/\dot \tau$.
An examination of the resulting damped oscillator equation shows that
diffusion suppresses the oscillation amplitude by a 
factor of order $e^{-k^2 \eta/ \dot\tau}$.
The damping scale $\kd$ is thus of order $\sqrt{\dot\tau/\eta}$,
corresponding to the geometric mean of the horizon and the mean free path.
Detailed numerical integration of the equations
of motion are required to track the rapid growth of the mean free path
and damping length through recombination itself.  
These calculations show that the damping 
scale is of order $\kd s_* \approx 10$ 
leading to a substantial suppression of the oscillations beyond the third peak
(see Fig.~\ref{fig:scheme}).  
Observations of the damping phenomena provide a check on the fundamental 
assumptions underlying the interpretation of the acoustic peaks.  
With $\Omega_b h^2$ and $\Omega_m h^2$ and the acoustic scale $l_A$ fixed
by the first few peaks, the damping scale is uniquely predicted given 
the atomic physics of recombination.  Any change in recombination, for example
due to a variation in the fine structure constant, will be revealed as a 
discrepancy in the predictions.  Likewise, any misinterpretation of $l_A$
due to say initial conditions that were not purely curvature fluctuations
resulting in a phase shift in the peaks would also show up as a discrepancy.
The data in the damping tail to date are beautifully consistent with the predictions
of the model (see Fig.~\ref{fig:data}).

\subsection{Polarization}
\label{sec:polarpeaks}

The dissipation of the acoustic oscillations leaves a signature in
the polarization of the CMB in its wake.
Recall that the observed polarization is a projection of the local quadrupole
temperature anisotropy at last scattering.
The fact that the polarization source is the quadrupole
explains the shape and height of the 
polarization spectra in Fig.~\ref{fig:data}.
Since the quadrupole is of order $kv_\gamma/\dot\tau \sim (k/k_{\rm d})(k_d \eta_*)^{-1} v_\gamma$ 
(see Eqn.~\ref{eqn:pitc}), 
the polarization
spectrum rises as $l/l_D$ to peak at the damping scale with an amplitude of
about 10\% of the temperature fluctuations before falling due to the elimination of
the acoustic source itself due to damping.
Since $v_\gamma$ is out of
phase with the temperature, the polarization peaks are also out of phase with the
temperature peaks. 
Furthermore, the phase relation also tells us that the polarization is 
correlated with the temperature perturbations.  The correlation power $C_l^{\Theta E}$ 
being the product of the two, exhibits oscillations at twice the acoustic frequency.
As in the case of the damping, the predicting the precise value requires
numerical work \citep{BonEfs87} 
since $\dot\tau$ changes so rapidly near recombination.  
Nonetheless the detailed predictions shown in Fig.~\ref{fig:scheme} bear these
qualitative features.

The acoustic polarization and cross correlation has recently been
detected by the DASI experiment \cite{Kovetal02}.  Like the 
damping scale, the acoustic polarization spectrum is uniquely predicted 
from the temperature spectrum once
$\Omega_b h^2$, $\Omega_m h^2$ are specified.   Polarization thus
represents a sharp test on the assumptions of 
the recombination physics and power law curvature fluctuations in 
the initial conditions used in interpreting the temperature peaks.

These acoustic peaks in the polarization appear exclusively in the 
$EE$ power spectrum due to the azimuthal symmetry of the plane wave fluctuations.
During the break down of tight coupling that occurs at last scattering,
any gravitational waves present will also imprint a local quadrupole anisotropy 
to the photons and hence a linear polarization to the CMB \cite{Pol85}.  
These contribute to
the $BB$ power and their detection would provide invaluable information on the
origin of the fluctuations (see e.g. \cite{SteTur01}).  
Specifically, in simple inflationary models their
amplitude gives the energy scale of inflation.  The gravitational wave amplitude $h$
oscillates and decays once inside the horizon, so the associated polarization source
scales as $\dot h /\dot \tau$ and so peaks at the $l \approx 100$
horizon scale and not the damping scale at recombination (see Fig.~\ref{fig:data}).
This provides a useful scale separation of the various polarization effects.

\section{Secondary Anisotropy}
\label{sec:secondary}

\subsection{Gravitational Secondaries}

As the photons travel from the last scattering surface to the observer, they traverse
the large-scale structure of the universe leaving 
subtle imprints in the temperature
and polarization fields.

There are two main gravitational effects.  Intervening
mass along the line of sight 
gravitationally lenses the CMB photon trajectories \cite{BlaSch87}
and hence distorts both the temperature
and polarization anisotropy fields.
The photons are deflected according to the angular gradient of
the potential projected along the line of sight,
\begin{equation}
\phi(\bn) = 2 \int_{\eta_*}^{\eta_{a=1}} 
d \eta\, {(D_*-D) \over D\, D_*} \Phi(D \bn,\eta)\,.
\end{equation}
Because surface brightness is conserved in lensing, the deflection
simply remaps the observed fields as
\begin{align}
X(\bn) & \rightarrow   X(\bn + \nabla\phi) \,,
\label{eqn:remap}
\end{align}
where $X \in  \{ \Theta$, $Q$, $U \}$.
The typical deflection angle is of order a few arcminutes but 
the lines of sight are coherently deflected across scales of
a few degrees.   Since the coherence scale of the acoustic features
is larger than the deflection
angle, the lensing effect can be calculated by Taylor expanding
Eqn.~(\ref{eqn:remap}) (e.g.~\cite{Hu00b}).  The result is a product of fields so that
in harmonic space the modes
are coupled to each other across a range $\Delta l \approx 60$  
set by the coherence of the deflection.
Heuristically, lensing distorts the hot and cold spots
formed by the acoustic oscillations and hence
the mapping of $k$ to $l$.

In the temperature power spectrum, this mode coupling 
smooths the acoustic peaks slightly \cite{Sel96b}.
For the polarization, the remapping not only smooths the $EE$-peaks
but actually generates $B$-mode
polarization \cite{ZalSel98}.
Remapping by the lenses preserves the orientation of
the polarization but warps its spatial distribution 
and hence does not preserve the
symmetry of the original $E$-mode.   
Gravitational lensing represents a fundamental obstacle
to detection of low amplitude gravitational waves from
inflation.

Because the lensed CMB distribution is not linear in the
fluctuations, higher order statistics
are promising probes 
of lensing effects \citep{Ber97,Ber98,ZalSel99,Zal00}.
In particular, the coupling of multipoles separated
by $\Delta l < 60$ can be used to construct
a minimum variance estimator of the deflection potential $\phi(\bn)$
out of pairs of moments \cite{Hu01b}.
These quadratic estimators 
are close in performance to the optimal but more complicated
maximum likelihood estimator \cite{HirSel02}.
Since small-scale polarization anisotropy is otherwise
free of cosmological $B$-modes, most of the signal-to-noise
in the reconstruction lies in the pairing of
$E$-modes to neighboring lensing-generated $B$-modes
\cite{HuOka02}.  This correlation can in principle be used
to map the deflection potential down to the $10'$ scale.  This mapping
can in turn be used to measure the
growth of structure from $z_*=10^{3}$ \cite{Hu01c} and hence the properties of the
dark matter, including any component from massive neutrinos.  It can
also be used to remove much of the contamination to the gravitational
wave $B$-modes and bring the detection threshold for the energy
scale of inflation down to a $E_i \approx {\rm few\,} 10^{15}$ GeV 
\cite{KnoSon02}.

Gravitational potentials can also change the temperature anisotropy
through gravitational redshifts.  Density perturbations cease to grow once
the dark energy dominates the expansion.
As in the case of the matter-radiation transition, the gravitational potentials must then 
decay.
The evolution of the gravitational potential under a smooth 
component of dark energy is
given by
\begin{eqnarray}
\frac{d^2 \Phi}{d \ln a^2}  +
\left[ \frac{5}{2} - \frac{3}{2} w_e(a) \Omega_{e}(a) \right]
\frac{d \Phi}{d \ln a}  +
\frac{3}{2}[1-w_e(a)]\Omega_{e}(a) \Phi =0\,.
\label{eqn:decay}
\end{eqnarray}
With the appropriate replacements, this equation also applies to 
smooth components of dark matter, e.g. a light neutrino.  A scalar
field candidate for dark energy is smooth out to the horizon scale
at any given time \cite{CalDavSte98}.   

As we have seen, a decay in the gravitational potential 
 causes an effective heating of the photons in a gravitational well.
Like the matter-radiation analogue, it
is intrinsically a large effect since the net change due to the decay 
is $5$ times the Sachs-Wolfe effect of $\Psi/3$. 
However since the opposite effect occurs in underdense regions,
the contributions are canceled as photons traverse many crests and troughs of the
potential perturbation during the matter-dark energy transition.
The effect, called the (late-time) integrated Sachs-Wolfe (ISW) effect \cite{SacWol67,KofSta85},
then appears only at large scales or low multipoles 
(see Fig.~\ref{fig:scheme}).  

The ISW effect 
is the most direct signature of the dark energy available in the CMB 
and is very sensitive to its equation of state and clustering properties
if it deviates far from 
a cosmological constant ($w_e > -1/2$, see Fig.~\ref{fig:cl}), unfortunately given
the large-scale nature of the effect, the limited number of samples of the low
order multipoles, or ``cosmic variance'', prevents a precise determination from 
the power spectrum.
A promising technique is to isolate the ISW effect through its cross correlation
with reconstructed lensing maps from the CMB \citep{GolSpe99,ZalSel99,Hu01c}.

\subsection{Scattering Secondaries}

The universe is observed to be ionized out to $z \approx 6$ and is
thought to have undergone reionization sometime between
between $7 \le z \simlt 30$ (see \cite{BarLoe01} for a review).
Consequently a minimum of a few percent of the photons have
rescattered since recombination  
(see Eqn.~\ref{eqn:mfp}).
The main effect from rescattering of the photons 
during reionization is a uniform suppression of the peaks by $e^{-\tau}$
as anisotropy is destroyed by the randomization of directions in scattering.
Since this can be confused with a change in the
initial amplitude of fluctuations, it is important to resolve this ambiguity
for the study of initial fluctuations and 
the growth rate of fluctuations, which itself probes the 
properties of the dark energy  and dark matter through Eqn.~(\ref{eqn:decay}).
Reionization is interesting in its own right since it can us about 
the sources of ionizing radiation from 
the first astrophysical objects formed in the universe.

The most promising means of isolating the reionization effect through
the CMB is from the 
generation of large scale polarization during recombination.
The rescattered radiation becomes polarized since
temperature inhomogeneities become anisotropies by projection
(see Eqn.~\ref{eqn:monopoleproj}), passing 
through quadrupole anisotropy when the perturbations are on the horizon 
scale at any given time ($k D_* \approx k \eta \approx l = 2$).  
The result is 
a peak in the $E$-mode power spectrum shown in Fig.~\ref{fig:data}.
In a perfect, foreground-free world, even the minimal signal is
within reach of the MAP \cite{MAP} and Planck \cite{Planck} satellites
and can be used to isolate the reionization epoch \citep{HogKaiRes82,Kapetal02}.

The dissipation of the acoustic oscillations at small scales also provides
a window through which to see subtler effects.  The main contribution 
beyond the damping tail comes from the rescattering of the CMB off of
hot electrons in clusters of galaxies, called the Sunyaev-Zeldovich effect
\cite{SunZel72}.  Inverse Compton scattering represents a net transfer of energy between 
the hot electron gas and the cooler CMB.  It leaves a spectral distortion in 
the CMB where photons on the Rayleigh-Jeans side are transferred to the Wien
tail.  The effect in individual clusters is now routinely measured and is
an important means of studying the physics of clusters and the dark energy
\cite{CarHolRee02}.  The effect of 
unresolved high redshift clusters may have already been detected in the 
CMB by the CBI and BIMA
experiments, albeit at an amplitude that is somewhat higher than expected
\cite{Bonetal02,KomSel02}.  Confirmation of the nature of this small scale 
excess will require measurements at multiple frequencies to test
its inverse Compton
nature.

There are a host of smaller effects due to Doppler shifts of the moving
ionized gas (see e.g.~\cite{HuDod02}).  
These present a significant challenge to detect and interpret
and are beyond the scope of this review. 

\section{Discussion}
\label{sec:discussion}

The tremendous experimental progress in CMB anisotropy studies over
the last few years has helped establish a standard paradigm for 
cosmology at intermediate epochs but has simultaneously raised
questions about the physics at the two opposite ends of time.
Simple inflationary models of the early universe
have so far passed the test of the
acoustic peaks.  They predict the near scale-invariant initial
curvature fluctuations in a spatially flat background that have now
been observed.  Moreover, spatial flatness appears to be maintained today 
by a mysterious form of missing or dark energy.

The first step into the future will come with the release of
data from the MAP satellite in early 2003 which in the full course of the
mission will provide the definitive
measurement of temperature anisotropy across the whole sky down to a quarter
of a degree. 
The MAP satellite will test the spectrum of
initial conditions beyond the simple constraints on the power law index 
which can be extracted today (e.g. \cite{WanSpeStr99,TegZal02}).  
These advances will come with the increased precision in
the temperature measurements and determination of the acoustic peaks in the
polarization and cross correlation.  
Measurement of the polarization peaks will not only eventually
double the amount of statistical information that can be 
extracted from the CMB but can also provide a sharp distinction between small changes
in the dynamics of the plasma, which affect temperature and polarization
differently, and small deviations from scale 
invariance of in  the initial conditions (e.g. \cite{EasGreKinShi02}),  
which affect the two alike.
Studies of the Gaussianity
of the acoustic fluctuations will also provide a strong test of
the inflationary model.
The increased precision on the matter-radiation ratio
will sharpen considerably the constraints on the dark energy.
MAP should also be able to detect or place limits on the 
amount of reionization in the universe through a large angle
bump in the polarization and hence remove a 
central ambiguity for dark energy and dark matter studies from
the peaks.  

The next generation of experiments dedicated to fine-scale, secondary
structure in the anisotropy will enable new tests of the dark energy
from structures in the universe such as galaxy clusters.
They will also provide a new handle on reionization through the Doppler
shift of CMB photons off of moving structures.

Further down the road lies the milestone of the gravitational lensing
of the peaks.  Detection will not only require more sensitive instruments
and higher angular resolution but also subtraction of galactic and extragalactic
foreground contaminants \cite{TegEisHudeO00,PruSetBou00,Bacetal02}.
In principle higher order statistics can be used to reconstruct the potential
field in projection with a large gains in sensitivity available if
the $B$-modes from lensing can not only be detected but accurately
mapped \cite{HuOka02}.  From this reconstruction the dark 
matter and dark energy dependent growth of structure can be measured.
For example gravitational lensing can enable tests of
the scalar-field hypothesis for the nature of the dark energy 
through cross-correlation with the integrated Sachs-Wolfe effect \cite{Hu01c}. 

The ultimate future milestone for the CMB is the detection of gravitational
waves from inflation through $B$-mode polarization.  Detection would
represent strong evidence for the inflationary model and pin down its
energy scale. The window of detectability can be extended to
energy scales above a few $10^{15}$ GeV but only if the $B$-modes from  
gravitational lensing are removed from a direct reconstruction.  

These future milestones will require much experimental effort to achieve
and much theoretical effort to interpret.  Nevertheless they provide
the hope that the next decade of studies of CMB temperature and polarization
anisotropy will be as fruitful as the last.

%%%%%%%%%%%%%%%%%%%%%%%%%%%%%%%%%%%%%%%%%%%%%%%%
%% BACKMATTER
%%%%%%%%%%%%%%%%%%%%%%%%%%%%%%%%%%%%%%%%%%%%%%%%

\begin{theacknowledgments}
WH is supported by NASA NAG5-10840 and the DOE OJI program
\end{theacknowledgments}

\bibliographystyle{aipproc}

\end{document}